# Multiplexed structured illumination super-resolution imaging with time-domain upconversion nanoparticles

Baolei Liu[1], Jiayan Liao[1], Yiliao Song[2], Jie Lu[2], Jiajia Zhou[1]*, Fan Wang[1,3]*

[1]Institute for Biomedical Materials and Devices (IBMD), Faculty of Science, University of Technology Sydney, NSW 2007, Australia
[2]Centre for Artificial Intelligence, Faculty of Engineering and IT, University of Technology Sydney, NSW 2007, Australia
[3]School of Electrical and Data Engineering, Faculty of Engineering and IT, University of Technology Sydney, NSW 2007, Australia

*e-mail: jiajia.zhou@uts.edu.au; fan.wang@uts.edu.au

**Abstract:** The emerging optical multiplexing within nanoscale shows super-capacity in encoding information by using the time-domain fingerprints from uniform nanoparticles. However, the optical diffraction limit compromises the decoding throughput and accuracy of the nanoparticles during wide-field imaging. This, in turn, challenges the quality of nanoparticles to afford the modulated excitation condition, and further to retain the multiplexed optical fingerprints for super-resolution multiplexing. Here we report a tailor-made time-domain super-resolution method with the lifetime-engineered upconversion nanoparticles for multiplexing. We demonstrate that the nanoparticles are bright, uniform, and stable under structured illumination, which supports a lateral resolution of 186 nm, less than 1/4th of excitation wavelength. We further develop a deep learning algorithm to coordinate with super-resolution images for more accurate decoding compared to a numeric algorithm. We demonstrate a three-channel sub-diffraction-limit imaging-based optical multiplexing with decoding accuracies above 93% for each channel, and larger than 60% accuracies for potential seven-channel multiplexing. The improved resolution provides high throughput by resolving the particles within the optical limit, which enables higher multiplexing capacity in space. This time-domain super-resolution multiplexing opens a new horizon for handling the growing amount of information content, diseases source, and security risk in modern society.

Optical multiplexing is an easily implemented solution that combines many signals onto an assemblage of optical carriers for information transmission [1,2], data storage [3,4], and security [5]. The relatively mature technique of wavelength or color division multiplexing has shown limitations in the requirement of multi-lasers for excitation and spectral cross-talk. The recently developed time-domain optical carriers have fundamentally resolved the spectral cross-talk issue under single laser excitation and subsequently enabled the super-capacity multiplexing in nanoscale dimension [6]. However, the sub-diffraction limit nanocarriers have not been able to fully exert their advantages, due to the restriction of optical diffraction limit during imaging. This will reduce the throughput of effective pixels within an imaging area. In particular, the spatial resolution further hampers the single nanocarriers for future digital assay in detecting low concentration analytes.

Towards the super-resolution multiplexing, organic fluorophores such as proteins and cyanine dyes have been explored for multi-color super-resolution imaging through DNA point accumulation in nanoscale topography (PAINT) and stochastic optical reconstruction microscopy (STORM) [7] and stimulated emission depletion (STED) [8] techniques. However, the existent issues from the color channel multiplexing remain in these experiments [9,10], e.g., the simultaneous employment of multi-color laser to match the depletion wavelength for different dyes, and the limited channel numbers by spectrum width to avoid cross-talk. As an orthogonal dimension to color, the time-domain can be used to boost the super-resolution capacity. This time-domain super-resolution imaging method technically challenges the quality of nanoprobes, which has to be bright, stable, and long lifetime under pulsed laser excitation.

In the search of the potential probe candidates, most of the nanoparticles such as quantum dots [11,12,13], carbon dots [14,15], organic dyes [16,17,18] show short lifetimes in nanoseconds, limiting the channels for time-domain multiplexing. The best candidate of upconversion nanoparticles (UCNPs) produces emissions from isolated lanthanide ions in a crystal field, in which the forbidden partially allowed f-f transitions show stable signals and an excited-state lifetime of a microsecond to millisecond range [6,19,20,21,22,23]. Moreover, the lifetime of a typical excited state can be arbitrarily tuned through the microlocal environment control within a nanocrystal, not to mention the diversity from diverse ions and excited states.

Here, we demonstrate that a library of UCNPs with tailored lifetime profiles show sufficient brightness and long-term photostability for time-domain super-resolution microscopy. Based on the lifetime profiles of UCNPs, we establish a time-resolved structured illumination microscopy (TR-SIM), enabling high-throughput multiplexing super-resolved imaging through a single 808 nm excitation laser. We demonstrate a three-channel super-resolution optical multiplexing with decoding accuracies above 93% and potential seven-channel multiplexing with more than 60% recognition accuracy, by optimizing the signal diversion process with a deep learning algorithm.

**Nanoparticles with time-domain optical tunability**

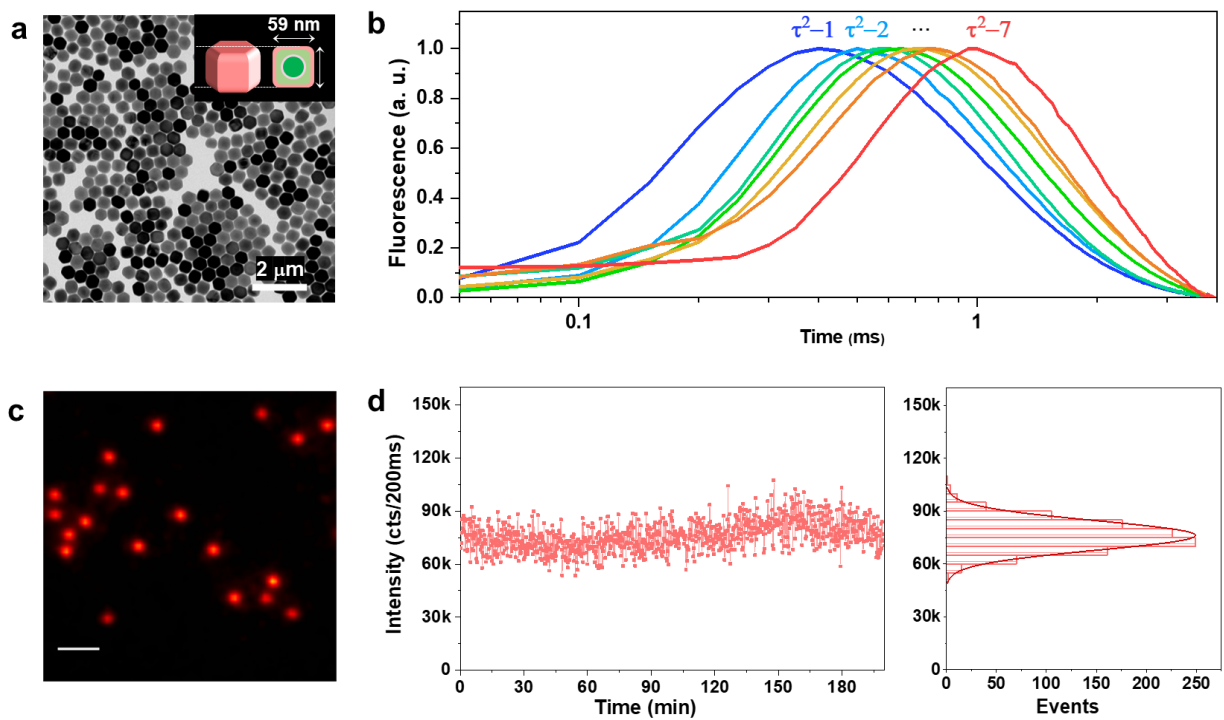

**Figure 1. Photo-stable uniform nanoprobes with time-domain optical tunability.** (a) TEM characterization of a kind of typical morphology uniform nanocrystals ($Yb^{3+}$-$Nd^{3+}$-$Er^{3+}$ core-multi shell nanoparticles) with an average size of 59 nm. An arbitrary batch (t²-5) of τ²-dot used here was chosen from a library of core-multi-shell nanoparticles we synthesized in this work. (b) Lifetime curves of τ²–x dots (x=1, 2, …, 7), with a rising time range from to 374μs to

710μs, a decreasing time range from 665μs to 1080μs, and peak shift from 616μs to 1147μs. (c) Widefield microscopic imaging of $\tau^2$-2 under 808 nm excitation. Scale bar: 2μm. (d) Photostability of a single dot upon 808 nm excitation in widefield microscopy.

We synthesized a library of UCNPs, which are sensitive to the 808 nm laser activation because of the $Nd^{3+}$ absorption, and $Nd^{3+}$ can pass the absorbed energy to $Yb^{3+}$, followed by multiphoton upconversion process to $Er^{3+}$, to produce visible emissions (See the Supplementary Materials and Methods; energy level diagram in Figure S5, and typical emission spectrum in Figure S7). TEM image in Figure 1a shows a typical $NaYF_4$ UCNPs with morphology uniformity, where the apertured architectures of each particle indicate a crystal phase of hexagonal and averaged size of about 60 nm in diameter. The inset illustration shows the design of the nanoparticles, including a core-multi-shell structure through layer-by-layer epitaxial growth (Supplementary Figure S1-S4). Following the previous works [6,23], the doping strategies to achieve lifetime tunability in plateau moment, rising time and decay time (named as $\tau^2$) in UCNPs' lifetime decay. We here made seven types of $\tau^2$-dots. From the averaged lifetime fingerprints of $\tau^2$-1 to $\tau^2$-7 (Figure 1b), the plateau moment shifts from 616 μs to 1147 μs, while the rising time and decay time prolong from 374μs to 710μs, and 664.8μs to 1080μs, respectively (Supplementary Figure S8).

Figure 1c shows an optical image of the $\tau^2$-2 dots under a wide-field microscopy. Within a widefield excitation (808nm, LU0808M250, Lumics Inc.), we observed uniform and bright UCNPs images where single nanoparticles can be distinguished from the cluster by evaluating its emission intensity. This high brightness allows for TR-SIM to see single UCNPs. The time-domain stability of $\tau^2$ dots is verified in Figure 1d, where the emission intensity from a single nanoparticle is stabilized at 75k counts per 200ms for more than 200 minutes, under 20% variation. Such excellent photostability is highly desired for super-resolution imaging when it requires long term and continuous recording.

**Fluorescence lifetime structured illumination microscopy**

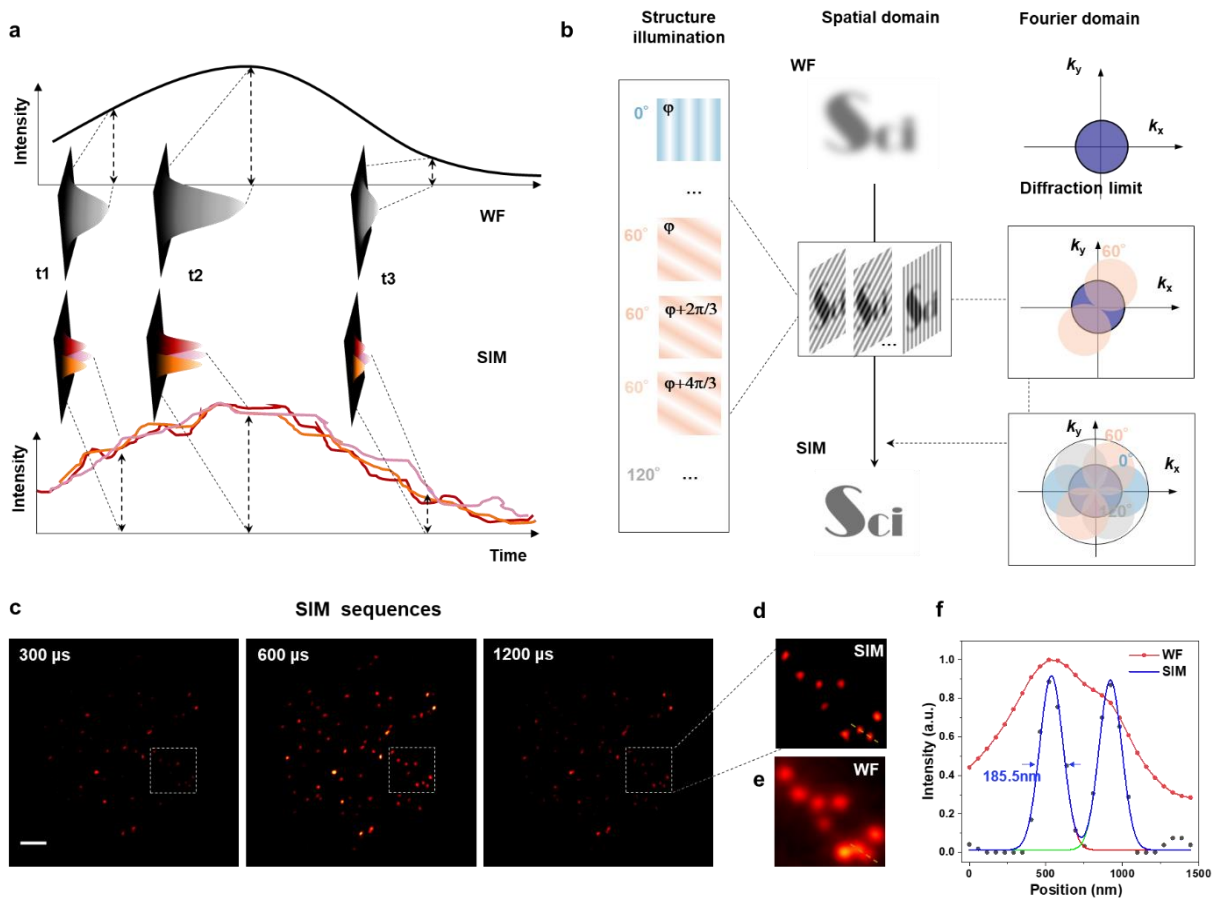

**Figure 2. Time-resolved structured illumination microscopy (TR-SIM) for sub-diffraction imaging.** (a) Illustration of the comparison of fluorescence lifetime imaging in widefield mode and SIM mode. (b)Principle of SIM method. Typically, 9 sinusoidal excitation patterns with differences in the orientations and phases are used in SIM. By exciting the sample with illumination patterns, the high-frequency super-resolution information can be shifted into the diffraction-limited detection passband. Thus, super-resolution images can be reconstructed with an extended spatial frequency spectrum. (c) Example of super-resolution imaging series in TR-SIM at 300 μs, 600 μs, and 1200 μs, which are selected from a detection range from 0 μs to 3000 μs with a time gate of 100 μs, under the laser excitation patterns of 0-200 μs. Scale bar: 2μm. (d) Enlarged imaging area under TR-SIM and corresponding area imaged by wide-field (WF) microscopy (e). (f) Line profile analysis of the selected area in (d).

Taking advantage of the $\tau^2$-dots in brightness, photostability, and microsecond scale lifetime, we developed a time-resolved super-resolution imaging method based on structured illumination microscopy [24,25,26,27,28,29,30] (TR-SIM, Supplementary Figure S6). Compared to traditional time-resolved wide-field fluorescence microscopy (TR-WF), the TR-SIM has superior decoding accuracy and much larger imaging throughput for optical multiplexing, benefiting from its ability to resolve the aggregated particles (Figure 2a) within diffraction limitation. Figure 2a is the diagram of TR-WF and TR-SIM. In TR-WF, under a wide-field Gaussian beam excitation, WF images at different time-sequence are collected by a time-resolved camera, and the evolution of integrated intensity shows the lifetime profile of a cluster of UCNPs. In TR-SIM, a series of excitation conditions are required (Figure 2b), and under each of the excitation, a group of time sequence images is taken with the time-resolved camera. The TR-SIM images are post-calculated from these series of images, with the ability to resolve single UCNPs from the cluster and show their individual lifetime profiles. In other words, for each of time sequence (e.g. t1, t2 or t3), the list of images with a series of excitation condition can be treated according to conventional SIM, as shown in Figure 2b. A diffraction limited sinusoidal pattern is used to expend the image collecting area in spatial frequency domain (Fourier transform of the images). The full enlarged area in spatial frequency domain requires three angles of patterns (e.g. 0°, 60° and 120°) to cover, and three phases of images for each angle to solve the reciprocal space patterns. Merging the 9 Fourier domain images (3 angles × 3 phases) and transferring it back to real-space domain, we are able to reconstruct the super-resolution image (Figure 2b). Figure 2c shows the typical TR-SIM images of $\tau^2$-2 dots at the time points of 300 μs, 600 μs and 900 μs from the pulse start point, with a time gate of 100 μs. The UCNPs can be clearly resolved by TR-SIM (Figure 2d), compared with WF image (Figure 2e). The line profile of a single UCNP (Figure 2f) indicates a lateral resolution of 186nm, 1/4$^{th}$ of excitation wavelength (808nm). We further show the examples of TR-SIM with seven kinds of $\tau^2$-2 dots in Figure 3. Compared with the WF imaging, the improved resolutions lead to higher numbers of single particles for a higher detection capacity. Note here each group of images are acquired from the peak time (range from 600 μs to 1200 μs) sequences in the lifetime profiles for better signal to noise ratio.

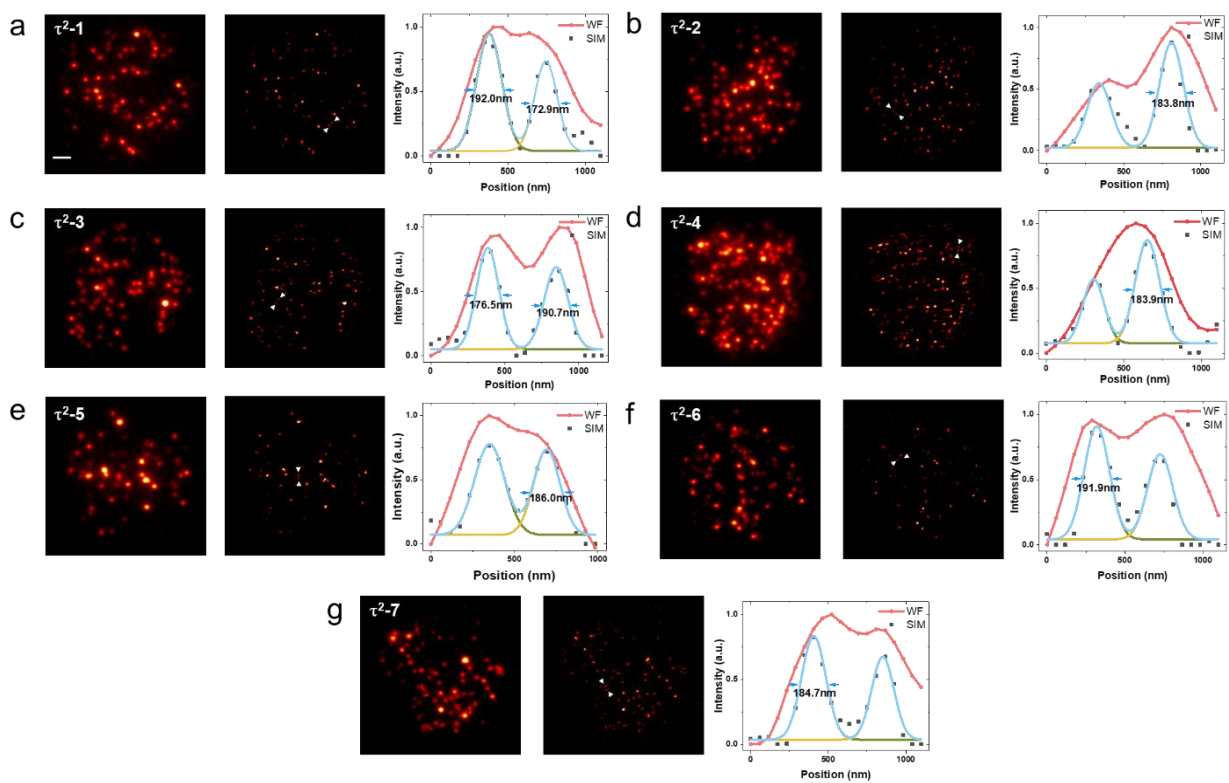

**Figure 3 Comparison of TR-WF and TR-SIM with seven kinds of τ2 dots.** For each group, TR-WF and TR-SIM and the profiles selected resolved particles (indicated by the arrows in TR-SIM images) are shown from left to right. Scale bar: 2μm.

In Figure 4, we developed a straightforward maximum correlation with mean lifetime curves (MC-MLC) method to use different $\tau^2$-dots as different channels for optical multiplexing. The working scheme of MC-MLC is shown in Figure 4a-c. The statistic lifetime profiles (Figure 4b) are averaged from many (typical 200) of single UCNP lifetime profiles (Figure 4a). With the database of averaged lifetime profiles for all seven types of dots, we can identify any unknown single dot profile by pointwise intensity comparison, in which the quantitative indicators (sum of squares of deviations $\sum_{x=1,2,...,30}(I_x - I_x^0)^2$ ) of each round comparison determines the category of the single dot (Figure 4c). Here $x$ represents the different time point, $I_x$ is the lifetime profile from unknow dot, and $I_x^0$ is the statistic lifetime profile from the known dots. Figure 4d shows the recognition results for $\tau^2$-1, $\tau^2$-4, and $\tau^2$-7 dots, in which the accuracies are higher than 90%. This proves the validity of our MC-MLC method. However, when we increase the categories of $\tau^2$-dots to six or seven types of dots, the accuracies of every dot downgrade with the lowest accuracy of 54.4% for $\tau^2$-2 (Figure 4e and 4f). The bowl-shaped recognition accuracies in Figure 4f indicate the challenge to identify the very similar lifetime profiles.

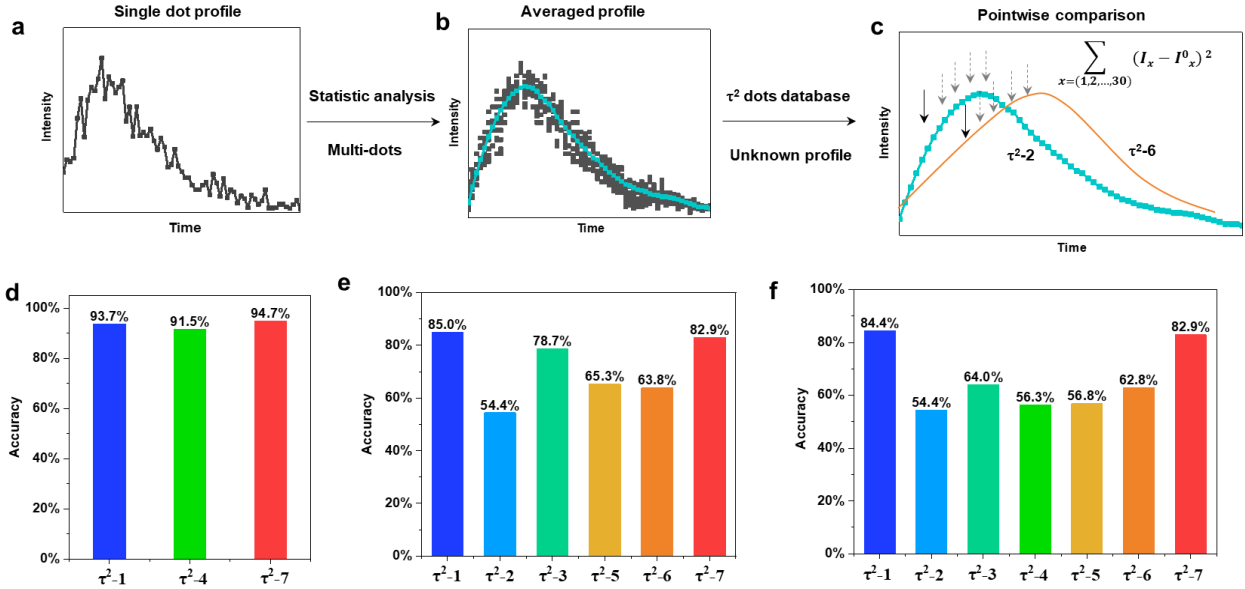

**Figure 4 Recognition of multiple single $\tau^2$-dot in SIM mode with MC-MLC method.** (a-c) Principle of MC-MLC method. The lifetime curve of a single dot (a) is compared with the mean curves of measured multiple $\tau^2$-dots (b). The least sum of squares of these deviations gives the classification accuracies (c). (d) Classification accuracies by maximum correlation with mean lifetime curves. (d–f) Classification accuracies of 3, 6, and 7 types of $\tau^2$-dots by MC-MLC method.

To improve multichannel recognition rates, we further developed a deep learning algorithm, as shown in Figure 5a. The algorithm is based on an architecture of convolutional neural networks (CNN) which contains a convolutional network and a fully connected network with two layers. We first apply the super-resolution image series of the seven types of single $\tau^2$-dots with a quantity of about 200 to 500 for training. The artificial neural network could extract the fingerprint features of each type of $\tau^2$-dots and provide the optimized CNN and classification boundaries by having feedback from accuracies. After the trained network was optimized, we then randomly selected a set of untrained image series for each type of dots (about 50 single dots for each type) to test and obtain their classification accuracies. To eliminate the random effect, we repeated 50 times validation experiment, in which 50 dots for each type are randomly selected for testing and the others are used for training and achieve the mean classification accuracies for each type of dots. For three-channel super-resolution multiplexing of $\tau^2$-1, 4, and 7, the mean classification accuracies of the lifetime profiles from selected single particles are further enhanced to 93%, 93%, and 96%, respectively (Figure 5b, Supplementary Figure S11). To demonstrate the power of the super-resolved optical multiplexing, we compared the recognition possibility of TR-SIM and TR-WF on the mixed samples of $\tau^2$-1, 4, and 7 in Figure 5c. Since the aggregation in WF mode is dramatically modifying the lifetime profile and affecting the sample recognition, TR-SIM helps to address this aggregation-induced artificial effect by separating single dots from the cluster (Figure 5c ii), benefiting from its higher lateral resolution (Figure 5c i). Moreover, the improved lateral resolution also helps to identify more particles (Figure 5c iii).

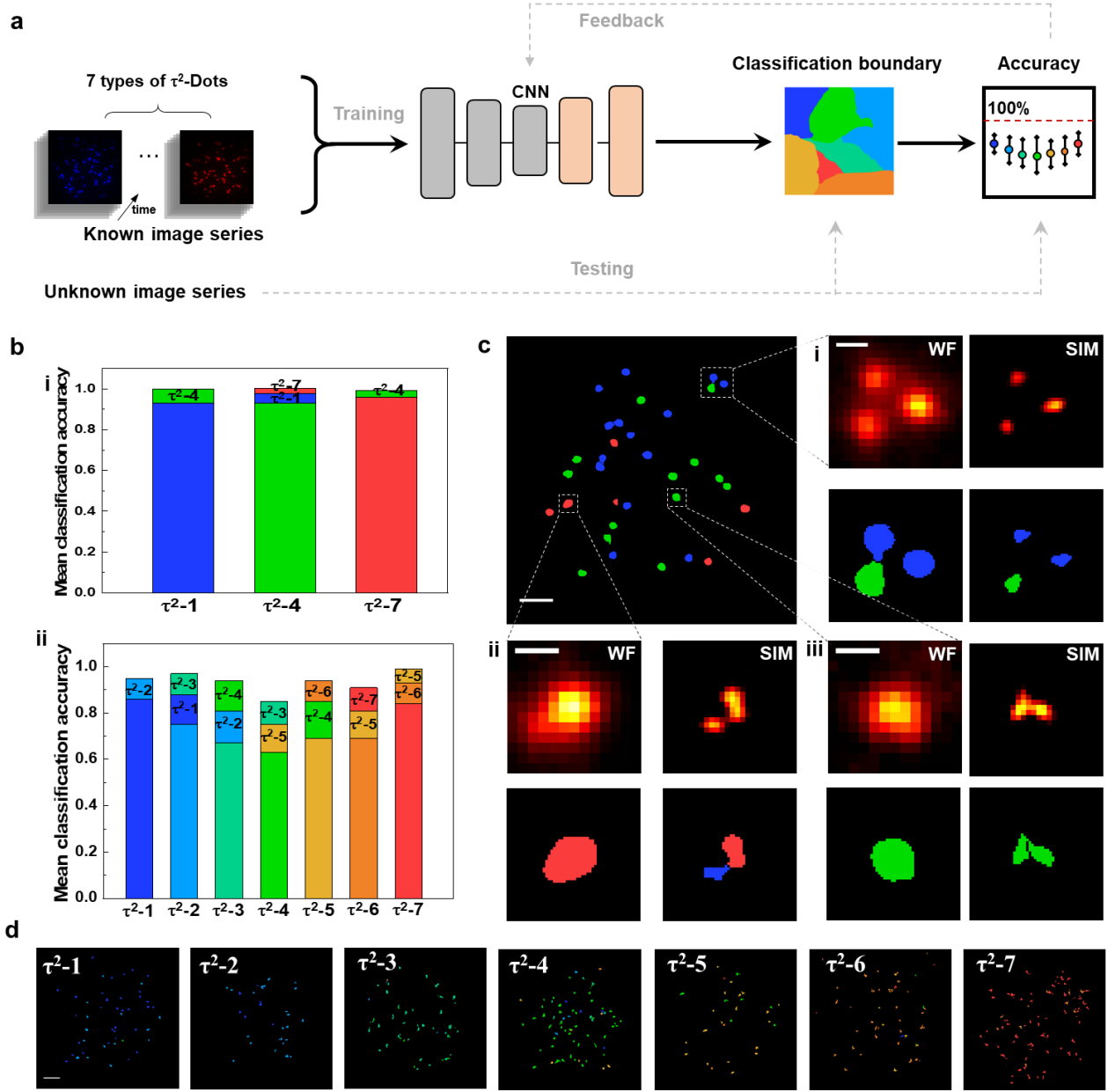

**Figure 5. Deep learning aided decoding of multiple single τ²-dots in super-resolution mode.** (a) Illustration of the flow path of deep learning aided decoding process. The time-resolved image series of seven types of single τ²-dot are input to convolutional neural networks (CNN) which contains a convolutional network and a fully connected network with two layers to define the feature coverage for each type of τ²-dots (the classification boundaries). Another randomly untrained seven sets of images are used as 'unknown dots' for the classification with acquired boundaries. Recognition accuracies are obtained by comparing recognition results and true categories of tested dots, based on 50 times random validation test of one set of seven types of dots. (b) Mean classification accuracies of 3 and 7 types of τ²-dots with deep learning algorithms. (c) Example of classification result for a mixture of three types of τ²-dots (τ²-1, 4, and 7 dots) in widefield mode (upper left). Scale bar: 2μm. i-iii: Imaging results (upper) and classification (bottom) of the selected small areas in WF mode and SIM mode show the precisely decoding and increased quantity of single dots due to the extended sub-diffraction limited resolution. Scale bar: 500nm. (d) Classification result for the well-trained CNN to recognize the arbitrary seven sets of 'unknown' τ²-dots (for visualization purpose, pseudo-color is used to represent each type of single dots corresponding to (b)). Scale bar: 2μm.

For the potential seven-channel super-resolution multiplexing, all the accuracies are higher than 60%, with a range from 63% to 86%, showing the better performance of deep learning algorithm than the

MC-MLC method, with a typical graphic classification result shown in Figure 5b i. Further assessing the unsuccessfully identified territories, we understand that the bias is mainly caused by the profile similarity from neighboring types of dots (Figure 5b & d). By removing one of the $\tau^2$-dots, e.g., $\tau^2$-4, the mean classification accuracies of all the six types of dots increased to higher than 70% (Supplementary Figure S9&10).

**Conclusion**

In conclusion, we have developed a super-resolution multiplexing imaging method with photostable time-domain nanoprobes: a library of $Yb^{3+}$-$Nd^{3+}$-$Er^{3+}$ core-multi shell UCNPs with tailored lifetime profiles. Rather than the implementation of multiple excitation lasers or separated detection, a simple optical setup contains a single excitation laser and a single camera is applied here for multichannel recognition of the probes. We show a three-channel super-resolution optical multiplexing with decoding accuracies above 93%, and potential seven-channel multiplexing with a 186 nm lateral resolution. Due to the nonlinear response of upconversion emission, this resolution can be further improved by the upconversion nonlinear SIM (U-NSIM) [30]. Comparing with the previous work of confocal based lifetime multiplexing of UCNPs-decorated microspheres [6] and WF decoding of $\tau^2$-dots [23], this work enhanced both imaging speed and lateral resolution for nanoscale multiplexing of single particles. We believe that time-domain nanoprobes will provide new insight into the multiplexed super-resolution method in the time dimension.

**Acknowledgments:** The authors acknowledge the financial support from the Australian Research Council Discovery Early Career Researcher Award Scheme (J. Z., DE180100669, F.W., DE200100074), China Scholarship Council Scholarships (Baolei Liu: No.201706020170, Jiayan Liao: No. 201508530231), National Natural Science Foundation of China (61729501), and Major International (Regional) Joint Research Project of NSFC (51720105015).

**Author contributions:** J.Z. and F.W. conceived the project and co-supervised the research; J.L. conducted synthesis, characterization, B.L. and J.L. collected data and processed the data; Y.S. conducted machine learning; B.L. and F.W. built the optical system; B.L., J.L., and J.Z. prepared the figures and Supplementary Materials; B.L., F.W. and J.Z. wrote the manuscript with input from other authors.

**Competing interests:** The authors declare no competing financial interest.